\documentstyle[aps,epsf]{revtex}
\twocolumn
\draft
\title{Magnetic Field Symmetry of Dynamic Capacitances}
\author{T. Christen and M. B\"uttiker}
\address{ D\'epartement de physique th\'eorique,
Universit\'e de Gen\`eve, 24 Quai Ernest-Ansermet \\ 
CH-1211 Gen\`eve, Switzerland}
\begin{document}
\maketitle
\begin{abstract}
While the static capacitance matrix is always symmetric and thus
an even function of the magnetic field,
the dynamic capacitance matrix of multi-terminal samples
obeys in general only the weaker Onsager-Casimir symmetry
relations. Our results are in accordance with
recent experimental observations of asymmetric dynamic capacitances
in quantized Hall systems. We predict quantization of the 
four-terminal resistances of an insulating Hall sample.  
\end{abstract}
%
%
%
\vspace{0.5cm}

Symmetry relations play an important role in both theoretical 
and experimental physics. In mesoscopic conductors, microreversibility 
coupled with the (irreversible) thermodynamics of electron reservoirs
leads to conductances which obey the Onsager-Casimir \cite{Cas} reciprocity
relations under magnetic field reversal. A manifestly symmetric formulation
of the dc-transmission approach to electrical conductance \cite{BUS,HBAS,DATTA} proved to be 
close to experimental reality \cite{BE86} and has been applied to a wide range 
of problems \cite{VANH} including the quantized Hall effect \cite{MBR}. 
In this Letter we are concerned not with stationary 
dc-transport but with the magnetic field symmetry of the dynamic 
capacitance matrix for arrangements of conductors which couple 
purely capacitively. In contrast to the dc-conductances, the static 
capacitance matrix can be 
obtained from thermodynamic potentials. As a consequence, the static capacitance 
matrix is symmetric and all elements of the capacitance matrix are even 
functions of the magnetic field. Clearly, as we increase the frequency the 
capacitances become themselves frequency dependent. A thermodynamic description
becomes invalid and the symmetry under magnetic field reversal might be weakened.
Below we show that the dynamic capacitance matrix is in general not an even function
of the magnetic field but obeys only the weaker Onsager-Casimir reciprocity symmetry.
This result is in agreement with a recent experiment by Sommerfeld et al. \cite{Som}.\\ \indent 
Linear electrical transport is characterized by a quadratic
admittance matrix $G_{\alpha \beta}$ with a dimension 
equal to the number of contacts which are
connected to the sample under investigation.
The Fourier amplitudes, $\delta I_{\alpha}$, of the current through 
contact $\alpha$ are then related to the Fourier amplitudes of the voltage
oscillation, $\delta V_{\beta} \exp(-i\omega t) $, at contact $\beta$ by
\begin{equation}
\delta I_{\alpha}=\sum_{\beta} G_{\alpha \beta}(\omega )\: \delta V_{\beta} \;\; . 
\label{eq1} 
\end{equation}
The Onsager-Casimir symmetry relations require that the admittance
matrix at a given magnetic field is equal to the transposed
admittance matrix at the reversed magnetic field:
\begin{equation}
(G_{\alpha \beta}(\omega))_{B}= (G_{\beta \alpha}(\omega))_{-B} \;\; .
\label{eq2} 
\end{equation}
The requirements of current conservation and of invariance against a global voltage shift
(gauge invariance) imply the additional sum rules $\sum _{\alpha }
G_{\alpha \beta} =\sum _{\beta } G_{\alpha \beta }=0$ \cite{BJPC}.
As a consequence, the admittance of a two-terminal sample
is determined by a single scalar admittance and is thus an even
function of the magnetic field $B$.  An asymmetric conductance, on the other hand,
requires at least a three-terminal sample.
While asymmetric dc-conductances are well established \cite{BE86,VANH},
the investigation of an asymmetric ac-response is relatively new.
Two years ago, Chen et al. \cite{Chen} reported on an asymmetric capacitance
of a gate close to the edge of a quantum Hall bar.
This effect was explained in the framework of the edge-state picture and
with the help of a novel theory of the low-frequency
admittance of mesoscopic conductors provided by Ref. \cite{BJPC}.
Importantly, this theory is based on the concept of partial densities of states
which are sensitive to the breaking of time-reversal symmetry.
In a succeeding work \cite{CB1} we have extended the theory to the general case of
a multi-terminal Hall sample with an arbitrary arrangement
of edge states. It is now understood that the capacitance
between two contacts is in general not an even function of the magnetic field,
if one of these contacts allows transmission of charge carriers to other
contacts. If, however, both contacts are purely capacitively coupled to the
rest of the sample and to each other, 
the static capacitance between them is always symmetric \cite{BJPC}. 
It would be tempting to extend this statement to the 
dynamical capacitance, i.e. to finite frequencies. 
However, in a recent experiment, Sommerfeld et al. \cite{Som} investigated 
the symmetry properties of gates which are capacitively
coupled to a disc-shaped two-dimensional electron gas. While in the
two-terminal case the expected symmetric capacitance was observed,
in the three-terminal case the capacitance turned out to be asymmetric.
In the present work we show that the above mentioned theory of
low-frequency admittance predicts the observed result. In particular, we
show that the dynamic capacitance
between two conductors which are both capacitively coupled to the rest of
the sample and to each other can be asymmetric, and we discuss an example similar to
the experiment of Sommerfeld et al. \cite{Som}.\\ \indent 

\begin{figure}[t]
 \epsfxsize = 80 mm
 \epsffile{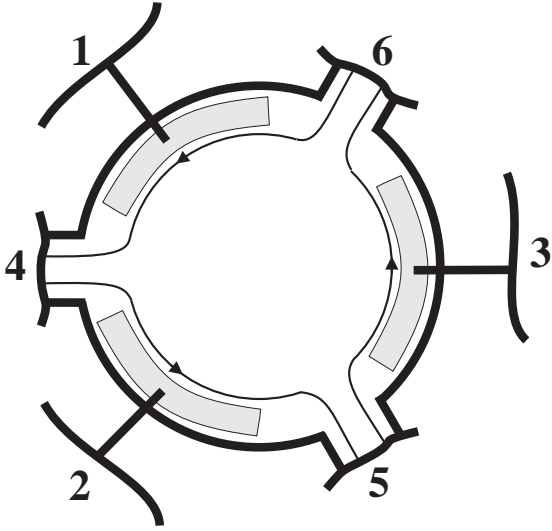}
\caption{ Disc-shaped 2-dimensional electron gas
at the first Landau level. Three gates ($1,2,3$) interact
electrically with the edge states. The fictitious contacts ($4,5,6$)
act as voltage probes and model dissipation.}
\label{fig}
\end{figure}

Consider a general multi-terminal sample consisting of conductors 
$j=1,...,M$ and which is connected via contacts
to electron reservoirs $\alpha = 1, ...,N$. For low frequencies we expand
the admittance in powers of frequency,  
\begin{equation}
G_{\alpha \beta} (\omega)= G_{\alpha \beta} ^{(0)}-i\omega E_{\alpha \beta}
+ \omega^{2} K_{\alpha \beta}  + {\cal O}(\omega ^{3}) \;\; ,
\label{eq3} 
\end{equation}
where $G_{\alpha \beta} ^{(0)} $ is the dc-conductance matrix,
$E_{\alpha \beta}$ is the emittance matrix \cite{BJPC}, and $K_{\alpha \beta}$ is related
to a charge relaxation time \cite{BJPC,BCW}.
If all conductors are purely capacitively coupled,
the dc-conductance vanishes and the emittance is
just the electrochemical capacitance which is a symmetric function
of the magnetic field \cite{BJPC}. 
In the sequel, we show that the matrix $K_{\alpha \beta}$ is in general
asymmetric even for purely capacitive contacts. To do this, we assume
the presence of a voltage probe at a contact $N$. Hence,
contact $N$ is connected to at least one other contact such that
$G_{N N}^{(0)}\neq 0$, and such that the net current through this contact vanishes,
$\delta I _{N }=0$. 
After an elimination of $\delta V _{N }$
from the Eqs. (\ref{eq1}) one obtains the reduced admittance matrix
$\tilde G_{\alpha \beta}$ for the remaining $N-1$ terminals. Now,
consider two purely capacitive contacts denoted by $\gamma$ and $\delta $.
(A specific example of such a structure is shown in row 3
of table \ref{tablesym}.) 
Thus, an element of the dc-conductance matrix $G_{\alpha \beta }^{(0)}$ vanishes
whenever one of the indices takes the value $\gamma $ or $\delta $.
An expansion of the reduced admittance with respect to frequency yields then the
reduced coefficients
\begin{eqnarray}
\tilde G_{\gamma \delta} & = & 0 \;\; , \label{eq3a} \\
\tilde E_{\gamma \delta} & = & E_{\gamma \delta} \;\; , \label{eq3b} \\
\tilde K_{\gamma \delta} & = & K_{\gamma \delta}+
\frac{E_{\gamma N}E_{N \delta}}{G_{N N}^{(0)}} 
\label{eq3c} \;\; .
\end{eqnarray}
Since contact $N $ is not purely capacitive, the emittance elements
$E_{\gamma N}$ and $E_{N \delta}$ are in general
asymmetric \cite{Chen,CB1}. Thus, $\tilde K_{\gamma \delta} $ is
in general asymmetric, too, even
for a symmetric $K_{\gamma \delta} $.\\ \indent
Starting from a $N$-terminal sample with a symmetric dynamic capacitance,
one can arrive at a sample with an asymmetric capacitance
for an $N-1$-terminal sample by using one
terminal as a voltage probe. Obviously, this is only possible if $N>3$
and provided that there are contacts with a non-vanishing dc-conductance.
This last condition was not satisfied in the experiment
by Sommerfeld et al. \cite{Som}, where apparently only capacitive contacts were present. 
Their sample, however, was macroscopic in size
such that electron motion is not phase coherent. But phase-breaking processes
(dissipation) can be modeled with the help of fictitious
voltage probes \cite{BUD}. We expect thus
an asymmetric dynamic capacitance of a three-terminal sample
in the presence of dissipation.\\ \indent
Consider a disc of a two-dimensional electron gas in a
perpendicular magnetic field as shown in Fig. \ref{fig}.
Three equivalent gates (grey regions) connected to the contacts ($1,2,3$) are
located at the sample boundary in a symmetric way and couple only capacitively. For simplicity, assume
that the electron gas is at the first Hall plateau such that one edge channel runs along the
sample boundary. Due to the fictitious voltage probes ($4,5,6$),
this edge channel is divided into three pieces which couple each to the closest gate
with a dynamic electrochemical capacitance of the form $C_{\mu}(\omega)=
C_{\mu}+i\omega (h/e^{2})C_{\mu}^{2}+{\cal O}(\omega ^{2})$ \cite{BJPC,BCW}, where
$C_{\mu}$ is the static electrochemical capacitance. Here, the dynamic correction term of the capacitance
can be associated with charge relaxation. The conductance of these channels is given by
$e^{2}/h$ for a spin split Landau level. The dc-relationship between currents and voltages
is obtained from the well-known transmission approach applied to quantized Hall systems \cite{MBR}. We
neglect all capacitances except those between gates and the closest piece of the edge channel. The
ac-part of the admittance can then be obtained as follows \cite{CB1,BCW}. A voltage oscillation
in reservoir $\beta $ induces a displacement current through
contact $\alpha $, if there is a capacitive coupling between the conductors in which contact $\beta $
injects charge, and the conductors which emit charge into contact $\alpha $.
For example, $\delta I_{4}\propto -i \omega (-C_{\mu}(\omega )\delta V_{1})$.
As usual, off-diagonal capacitance elements occur with a minus sign.
If a contact $\beta $ transmits carriers into contact $\alpha$, the associated capacitance element
is obtained from current conservation, e.g.,
$\delta I_{4}\propto -i \omega C_{\mu}(\omega )\delta V_{6}$.
In the purely capacitive case this capacitance element corresponds to a diagonal element of the capacitance,
e.g. $\delta I_{1}\propto -i \omega C_{\mu}(\omega )\delta V_{1}$. This receipt yields 
\begin{eqnarray}
\delta I_{1} & =&  -i\omega C_{\mu}(\omega ) (\delta V_{1}-\delta V_{6}) \nonumber \\
\delta I_{2} & = & -i\omega C_{\mu} (\omega )(\delta V_{2}-\delta V_{4}) \nonumber \\
\delta I_{3} & = & -i\omega C_{\mu} (\omega )(\delta V_{3}-\delta V_{5}) \nonumber \\
\delta I_{4} & =& (e^{2}/h)(\delta V_{4}-\delta V_{6}) -i\omega C_{\mu} (\omega )(\delta V_{6}-\delta V_{1}) 
\nonumber \\
\delta I_{5} & =  & (e^{2}/h)(\delta V_{5}-\delta V_{4}) - i\omega C_{\mu}(\omega ) (\delta V_{4}-\delta V_{2}) 
\nonumber \\
\delta I_{6} & = & (e^{2}/h)(\delta V_{6}-\delta V_{5})  - i\omega C_{\mu} (\omega )(\delta V_{5}-\delta V_{3}) 
\label{eq} \;\; .
\end{eqnarray}
Using the conditions for voltage probes, $\delta I_{4} = \delta I_{5}=
\delta I_{6}=0$, the elimination of $\delta V_{4}$, $\delta V_{5}$,
and $\delta V_{6}$ leads to the reduced admittance matrix $\tilde G_{\alpha \beta}$
for the contacts $1,2,3$. We find 
\begin{eqnarray}
\tilde G_{12} =   \frac{1}{3}\: (i\omega C_{\mu}-(h/e^{2})\omega ^{2}C_{\mu}^{2}) 
\label{eq7} \\
 \tilde G_{21} =   \frac{1}{3}\: i\omega C_{\mu}
\label{eq8} \;\; .
\end{eqnarray}
The other elements of $\tilde G_{\alpha \beta}$ are determined by the above mentioned
sum rules and by the specific symmetry of the sample (i.e. $\tilde G_{1 2}= \tilde G_{23}=\tilde G_{31}$
and $\tilde G_{21}= \tilde G_{13}=\tilde G_{32}$). There is obviously an asymmetry
$(\tilde G_{12})_{B}-(\tilde G_{12})_{-B}=-(h/3e^{2})\omega ^{2}C_{\mu}^{2}$ in ${\cal O} (\omega ^{2})$.
For the modulus it holds $|\;( \tilde G_{12})_{B}\; |>|\; (\tilde G_{12})_{-B}\; |$,
which is in accordance with the experiment \cite{Som}.
An additional elimination of contact $3$ leads to a (scalar and thus symmetric)
two-terminal capacitance equal to $3/2$ times the capacitance
associated with $\tilde G_{12}$ of Eq. (\ref{eq7}). This discussion suggests that dissipation plays 
an essential role. It would be interesting to investigate the asymmetry 
in the capacitance elements as a function of temperature or for models
in which the dissipation can be gradually switched off.\\ \indent
Instead of the just considered case where three contacts are capacitively coupled to
three pieces of an edge channel, we mention  the
analogous case with four gates coupled to four pieces of an edge
channel. For such a sample one can calculate the longitudinal resistance, $R_{L}$, and
the quantum Hall resistance, $R_{H}$ \cite{MBR}. Interestingly, we find {\em exact} quantization,
$R_{H}=h/e^{2}$ and $R_{L}=0$ independent of $\omega$.
Note that, due to the purely capacitive coupling of the reservoirs, this
sample is insulating, i.e. the dc resistances diverge. Clearly, for the four-terminal analog of Fig. 1 such a behavior is 
to be expected: Four probe measurements are independent of the contact impedances.
Furthermore, if the sample exhibits inelastic relaxation and equilibration
the manner in which current is introduced into the sample is unimportant.
We remark that the quantization of the four-terminal resistances is {\em not} a 
consequence of the symmetry of the model of Fig. 1 and is valid for arbitrary
capacitances between the gates and the closest edge. However, the absence  
of capacitive coupling along the edge and across the sample is crucial. 
In general, there are displacement current contributions and relaxation contributions
to both the Hall resistance and the longitudinal resistance\cite{CB1,BCW}.
Finally, we mention that even so the purely capacitively coupled sample of Fig. 1 is 
an insulator in terms of a strict dc-measurement, it is not a model of a 
Hall insulating state\cite{HS} for which it is expected 
that the Hall resistance remains finite 
but the longitudinal resistance diverges. We conclude with the remark 
that measurements on such samples are made with ohmic contacts, but that 
these contacts provide increasingly capacitive coupling only when the sample is 
driven deep enough into the insulating phase.
To conclude, we summarize the main results on the symmetry properties
of capacitances in table \ref{tablesym} for
two- and three-terminal samples. In order to
discuss the symmetry properties of the admittance under reversal of a magnetic field
one must first distinguish the leading order response ($E_{\alpha \beta}$,
proportional to $\omega$) and the higher 
order response ($K_{\alpha \beta}$, proportional to $\omega^{2}$, or even
higher powers of frequency). Secondly,
one must distinguish between different conducting topologies. The static capacitance is
always symmetric since it is the second derivative of an appropriate thermodynamic potential.
The dynamic capacitance of two-terminal samples (row 1 of table \ref{tablesym})
is also always symmetric.
This is a direct consequence of the Onsager-Casimir
relations, gauge invariance, and current conservation. Purely capacitive, phase-coherent
multi-terminal samples (row 2 of table \ref{tablesym})
exhibit dynamic capacitance coefficients which are even functions of
the magnetic field.
In the case of a spatially dependent potential along a conductor (e.g.
in the presence of a voltage probe), the emittance of capacitive contacts is symmetric
but higher-order terms in frequency are not (row 3 of table \ref{tablesym}).
In a measurement for which at least one of the
conductors is connected to two or more contacts (and thus permits transmission from one of
its contacts to another) one finds an emittance\cite{Chen}
which is not an even function of the magnetic field (row 4 of table \ref{tablesym}).

Table \ref{tablesym}) demonstrates that dynamic capacitances have a far richer 
magnetic field symmetry behaviour than dc conductances. 

{\em Acknowledgment -}
This work was supported by the Swiss National Science Foundation under grant
Nr. 43966.

\newpage

%
%
\begin{table}
\centering\ifx\sptloaded\undefined\small\else\fi
\caption[ ]{Symmetry properties of the dynamic
capacitance of two- and three-terminal
samples; a circle ($V$)  indicates a voltage probe.}
\hspace{1mm}
\vspace{-1cm}
 \epsfxsize = 70 mm
 \epsffile{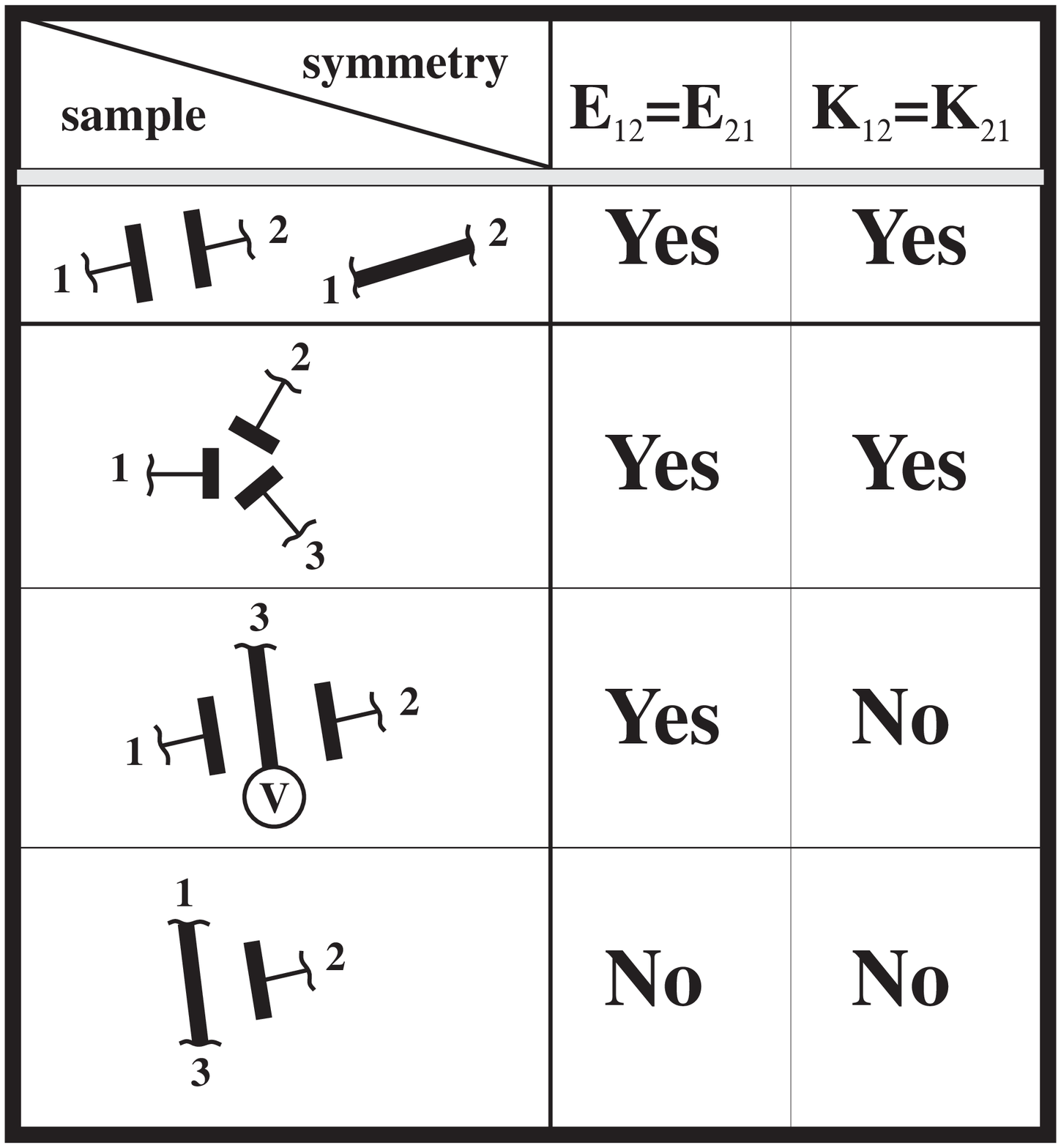}
\label{tablesym}
\end{table}
%
%
\end{document}